\documentclass[11pt,onecolumn]{article}
\usepackage[utf8]{inputenc}
\usepackage[english]{babel}
\usepackage[numbers]{natbib}
\usepackage[top=2cm, bottom=2cm, left=2cm,right=2cm]{geometry}

\usepackage{graphicx}%
\usepackage{xurl}
\usepackage[ruled]{algorithm2e}
\usepackage{algpseudocode}
\graphicspath{{./figures/}}
\usepackage{multirow}%
\usepackage{amsmath,amssymb,amsfonts}%
\usepackage{amsthm}%
\usepackage{mathrsfs}%
\usepackage[title]{appendix}%
\usepackage{xcolor}%
\usepackage{textcomp}%
\usepackage{manyfoot}%
\usepackage{booktabs}%
\usepackage{listings}%
\usepackage{tikz,lipsum,lmodern}
\usetikzlibrary{matrix,arrows.meta, positioning, fit, backgrounds}
\usepackage{amsmath,amssymb,amsthm}

\theoremstyle{definition}

\usepackage{enumerate}
\usepackage[braket, qm]{qcircuit}

\usepackage{caption}
\usepackage{subcaption}
\usepackage[most]{tcolorbox}

\newcommand{\braket}[2]{\ensuremath{\left\langle1|2\right\rangle}} 
\newcommand{\wmc}[1]{\ensuremath{\widetilde{\mathcal{1}}}}
\renewcommand{\bf}[1]{\ensuremath{\mathbf{1}}}
\newcommand{\norm}[1]{\ensuremath{\lVert1\rVert }}
\newcommand{\revfix}[1]{{1}}
\let\vec\mathbf

\title{Quantum  Simulations Based on Parameterized Circuit of an Antisymmetric Matrix}
\author{Ammar Daskin\thanks{adaskin25@gmail.com}}

\date{}
\begin{document}
\maketitle
\begin{abstract}
Given an antisymmetric matrix $A$ or the unitary matrix $U_A = e^A$—or an oracle whose answers can be used to infer information about $A$—in this paper we present a parameterized circuit framework that can be used to approximate a quantum circuit for $e^A$.  
We design the circuit based on a uniform antisymmetric matrix with $\{\pm 1\}$ elements, which has an eigenbasis that is a phase-shifted version of the quantum Fourier transform, and its eigenspectrum can be constructed by using rotation $Z$ gates. 
Therefore, we show that it can be used to directly estimate $e^A$ and its quantum circuit representation. Since the circuit is based on $O(n^2)$ quantum gates, which form the eigendecomposition of $e^A$ with separate building blocks, it can also be used to approximate the eigenvalues of $A$.  
\end{abstract}

\section{Introduction}  
The time evolution operator—the solution to the Schrödinger equation—is given by the unitary operator \( U(t) = e^{it\mathcal{H}} \), where \( \mathcal{H} \) is the Hamiltonian whose eigenvalues correspond to the quantized energy levels and are central to understanding the properties of a quantum system. Thus, simulating quantum systems often reduces to computing the eigenspectrum of the Hamiltonian matrix \cite{lloyd1996universal}.  
On general-purpose quantum computers, which process information through unitary operations (quantum gates), such simulations require approximating \( U(t) = e^{it\mathcal{H}} \) as sequences of elementary quantum gates, also represented by unitary matrices \cite{nielsen2010quantum}.  
Alternatively, there are quantum algorithms employing block encoding techniques \cite{daskin2012universal,low2019hamiltonian,gilyen2019quantum}, where \( \mathcal{H} \) is embedded as a subcomponent of a larger unitary matrix. In such schemes, an ancillary register is introduced with a control qubit to condition the action of the unitary matrix. Measuring specific outcomes of this ancillary register, the eigenspectrum of \( \mathcal{H} \) can be directly resolved.  
While these methods help to bypass explicit gate-level decomposition of \( U(t) \) and enable direct spectral analysis, the success probability of the condition on the ancillary register and the difficulty of defining the behavior in the case of consecutive applications of the unitary make them challenging to apply in practical problems.  

Parameterized quantum circuits \cite{daskin2012universal} have been one of the main tools in recent advances in quantum computing. 
These circuits incorporate adjustable parameters (e.g., \cite{peruzzo2014variational, romero2017quantum}) to iteratively optimize quantum operations, offering a classical-quantum hybrid framework for simulating quantum Hamiltonians (e.g., \cite{bauer2016hybrid}). This adaptive framework through parameterized quantum circuits with the help of a classical optimization has expanded their utility beyond Hamiltonian simulation to solving optimization \cite{tilly2022variational} and machine learning problems \cite{benedetti2019parameterized}, bridging the gap between theoretical quantum algorithms and practical implementation challenges \cite{cerezo2022challenges,combarro2023practical}.

\subsection{Motivation for Considering Antisymmetric Matrices}  

Skew-symmetric (or antisymmetric) matrices appear in various fields: They are the primary tool for representing directed (or oriented) graphs, and their spectral properties—such as graph energy \cite{adiga2010skew}—are used in applications like graph clustering \cite{hayashi2022skew} and chemistry \cite{cavers2012skew,li2013survey}.  
Moreover, low-rank approximations of antisymmetric matrices are employed in quantum chemistry \cite{kovac2017structure,ceruti2020time}, and their antisymmetric properties arise in classical and quantum machine learning algorithms \cite{klus2021symmetric}.  

The Hamiltonian matrix $\mathcal{H}$ governing the dynamics of a quantum system is generally considered to be complex Hermitian to ensure that its eigenvalues—representing the system's energy values—are real.  
A complex Hermitian matrix $C$ can be expressed in terms of real symmetric ($S$) and antisymmetric ($A$) matrices as $C = S + iA \in \mathbb{C}$. For vectors $x = u + iv$ and $y = p + iq$ with $u, v, p, q \in \mathbb{R}$, instead of solving $Cx = y$ directly, the following real-valued system can be used to avoid complex arithmetic \cite{trabelsi2017deep,bassey2021survey,lee2022complex}:

\begin{equation}
    M \begin{pmatrix} u \\ v \end{pmatrix} = \begin{pmatrix} p \\ q \end{pmatrix}, \quad \text{where} \quad M = \begin{pmatrix} S & -A \\ A & S \end{pmatrix}.
\end{equation}

The matrix $M$ is a fundamental structure for the real representation of complex matrices, enabling the application of iterative methods to solve complex-valued linear systems (e.g., \cite{axelsson2000real,benzi2008block,day2001solving}). This approach allows the use of well-established iterative solvers such as GMRES \cite{saad1986gmres}, BiCGSTAB \cite{gutknecht1993variants}, and the conjugate gradient method \cite{freund1992conjugate} without requiring complex arithmetic support in the solver implementation.  

This representation also facilitates eigen-decomposition, as the resulting matrix $M$ retains the original eigenvalues (with multiplicity). Additionally, complex-valued neural networks in machine learning employ a similar decomposition, representing complex weights as $W = W_S + iW_A$ in a form analogous to $M$. This enables forward passes and gradient calculations using standard real-valued matrix operations \cite{trabelsi2017deep,bassey2021survey,lee2022complex}. Thus, the matrix structure $M$ is widely used to transform complex linear algebra problems into equivalent real-valued formulations.  

A recent paper \cite{hoffreumon2025quantum} proposed using the same complex representation in quantum systems by employing a symmetric block representation of the complex Hermitian Hamiltonian and replacing the Kronecker product $\otimes$ with a new combination map $\odot$.  
For the Hermitian Hamiltonian matrix $\mathcal{H} = S + iA$, where $S$ and $A$ are its symmetric and antisymmetric parts, the introduced combination map $\odot$ (Ref. \cite{hoffreumon2025quantum}) preserves the real representation when combining subsystems. This enables the real representation to serve as an extension for quantum systems.  

The unitary evolution of the Hamiltonian, constructed via the matrix extension $M$, can be approximated using Trotter-Suzuki decomposition—particularly for $U_A = e^A$. For example, the exponential of the Hamiltonian can be approximated as:  
\begin{equation}
    e^{-i\mathcal{H}} \approx \left(e^{-iS/t}e^{A/t}\right)^t.
\end{equation}  
Thus, an accurate approximation of this unitary provides a pathway to simulate the Hamiltonian when the circuit for the symmetric part $S$ is known.  

Simulating a matrix $A$ requires decomposing $e^{A} \approx U_1 U_2 \dots U_K$, where each $U_i$ ($0 \leq i \leq K$) corresponds to an implementable quantum gate (or a known unitary operation). Here, $K$ determines the computational complexity of the circuit, which directly impacts the efficiency of algorithms relying on $e^A$. Consequently, developing tools to design circuit approximations for $e^A$ is crucial as a compiler component for problems involving antisymmetric matrices $A$.  

\subsection{Main Contribution}

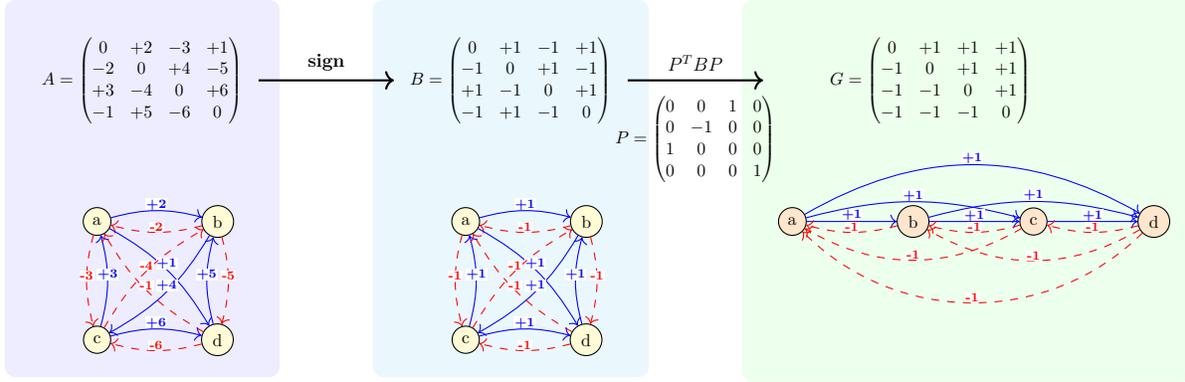
\begin{figure}[ht]
    \centering
    \usetikzlibrary{arrows.meta, positioning, fit, backgrounds}

\begin{tikzpicture}[
    node distance=2cm,
    edge/.style={->},
    posedge/.style={edge, blue, solid},
    negedge/.style={edge, red, dashed},
    label/.style={font=\scriptsize\bfseries, fill=white, inner sep=0.5pt},
    op/.style={font=\large\bfseries, above, inner sep=6pt},
    matrix node/.style={inner sep=10pt},
    column bg/.style args={#1/#2}{%
        rectangle,
        rounded corners,
        inner sep=8pt,
        fill=#1!10,
        opacity=0.7,
        fit={#2}
    }
]

\begin{scope}[scale=0.6, every node/.style={transform shape}]

\node[matrix node] (A) {
    $A = \begin{pmatrix}
    0 & +2 & -3 & +1 \\
    -2 & 0 & +4 & -5 \\
    +3 & -4 & 0 & +6 \\
    -1 & +5 & -6 & 0
    \end{pmatrix}$
};

\node[circle, draw, fill=yellow!20, below=1.5cm of A, xshift=-1cm] (a) {a};
\node[circle, draw, fill=yellow!20, right=of a] (b) {b};
\node[circle, draw, fill=yellow!20, below=of a] (c) {c};
\node[circle, draw, fill=yellow!20, right=of c] (d) {d};

\draw[posedge] (a) to[bend left=15] node[label, above] {+2} (b);
\draw[negedge] (b) to[bend left=15] node[label, above] {-2} (a);
\draw[negedge] (a) to[bend right=15] node[label, above] {-3} (c);
\draw[posedge] (c) to[bend right=15] node[label, above] {+3} (a);
\draw[posedge] (a) to[bend left=15] node[label, above] {+1} (d);
\draw[negedge] (d) to[bend left=15] node[label, above] {-1} (a);
\draw[posedge] (b) to[bend left=15] node[label, above] {+4} (c);
\draw[negedge] (c) to[bend left=15] node[label, above] {-4} (b);
\draw[negedge] (b) to[bend left=15] node[label, above] {-5} (d);
\draw[posedge] (d) to[bend left=15] node[label, above] {+5} (b);
\draw[posedge] (c) to[bend left=15] node[label, above] {+6} (d);
\draw[negedge] (d) to[bend left=15] node[label, above] {-6} (c);
\node[matrix node, right=3cm of A] (B) {
    $B = \begin{pmatrix}
    0 & +1 & -1 & +1 \\
    -1 & 0 & +1 & -1 \\
    +1 & -1 & 0 & +1 \\
    -1 & +1 & -1 & 0
    \end{pmatrix}$
};

\draw[->, thick] 
    (A.east) -- node[op] {sign} (B.west);

\node[circle, draw, fill=yellow!20, below=1.5cm of B, xshift=-1cm] (a2) {a};
\node[circle, draw, fill=yellow!20, right=of a2] (b2) {b};
\node[circle, draw, fill=yellow!20, below=of a2] (c2) {c};
\node[circle, draw, fill=yellow!20, right=of c2] (d2) {d};

\draw[posedge] (a2) to[bend left=15] node[label, above] {+1} (b2);
\draw[negedge] (b2) to[bend left=15] node[label, above] {-1} (a2);
\draw[negedge] (a2) to[bend right=15] node[label, above] {-1} (c2);
\draw[posedge] (c2) to[bend right=15] node[label, above] {+1} (a2);
\draw[posedge] (a2) to[bend left=15] node[label, above] {+1} (d2);
\draw[negedge] (d2) to[bend left=15] node[label, above] {-1} (a2);
\draw[posedge] (b2) to[bend left=15] node[label, above] {+1} (c2);
\draw[negedge] (c2) to[bend left=15] node[label, above] {-1} (b2);
\draw[negedge] (b2) to[bend left=15] node[label, above] {-1} (d2);
\draw[posedge] (d2) to[bend left=15] node[label, above] {+1} (b2);
\draw[posedge] (c2) to[bend left=15] node[label, above] {+1} (d2);
\draw[negedge] (d2) to[bend left=15] node[label, above] {-1} (c2);

\node[matrix node, right=3cm of B] (C) {\hspace{1cm}
    $G = \begin{pmatrix}
    0 & +1 & +1 & +1 \\
    -1 & 0 & +1 & +1 \\
    -1 & -1 & 0 & +1 \\
    -1 & -1 & -1 & 0
    \end{pmatrix}$
};

\draw[->, thick] 
    (B.east) -- node[op](op2) {$P^{T}BP$} (C.west);

\node[matrix node, below=-0.01cm of op2] (new) {$P = \begin{pmatrix} 0 & 0 & 1 & 0 \\ 0 & -1 & 0 & 0 \\ 1 & 0 & 0 & 0 \\ 0 & 0 & 0 & 1 \end{pmatrix}$};
    
\node[circle, draw, fill=orange!20, below=1.5cm of C, xshift=-2.5cm] (a3) {a};
\node[circle, draw, fill=orange!20, right=of a3] (b3) {b};
\node[circle, draw, fill=orange!20, right=of b3] (c3) {c};
\node[circle, draw, fill=orange!20, right=of c3] (d3) {d};

\draw[posedge] (a3) -- node[label, above] {+1} (b3);
\draw[posedge] (a3) to[bend left=15] node[label, above] {+1} (c3);
\draw[posedge] (a3) to[bend left=30] node[label, above] {+1} (d3);
\draw[posedge] (b3) -- node[label, above] {+1} (c3);
\draw[posedge] (b3) to[bend left=15] node[label, above] {+1} (d3);
\draw[posedge] (c3) -- node[label, above] {+1} (d3);

\draw[negedge] (b3) to[bend left=15] node[label, above] {-1} (a3);
\draw[negedge] (c3) to[bend left=30] node[label, above] {-1} (a3);
\draw[negedge] (d3) to[bend left=45] node[label, above] {-1} (a3);
\draw[negedge] (c3) to[bend left=15] node[label, above] {-1} (b3);
\draw[negedge] (d3) to[bend left=30] node[label, above] {-1} (b3);
\draw[negedge] (d3) to[bend left=15] node[label, above] {-1} (c3);

\node[below=2.5cm of c3, xshift=-2cm] (bnode) {};
\begin{scope}[on background layer]
    \node[column bg=blue/{(A) (a) (b) (c) (d)}] {};
    \node[column bg=cyan/{(B) (a2) (b2) (c2) (d2)}] {};
    \node[column bg=green/{(C) (a3) (b3) (c3) (d3) (bnode)}] {};
\end{scope}

\end{scope}
\end{tikzpicture}
    \caption{
        The intuition behind using a quantum circuit for $G$ to simulate an antisymmetric matrix. 
        If $B$ is acyclic (like $G$), it can be converted into $G$ using a permutation matrix $P$. 
        If $B$ is cyclic, it can be transformed into $G$ via a signed permutation matrix with $\pm 1$ elements.
        The approach simulates $A$ (i.e., $e^A$) using parameterized circuits for $e^G$ and $P$.
    }
    \label{fig:Intuition}
\end{figure}

Given an antisymmetric matrix $A$ (or its unitary counterpart $U_A = e^A$)—or an oracle providing information about $A$—our main contribution is a parameterized circuit framework to approximate quantum circuits for $e^A$.  

We design the circuit by first analyzing an antisymmetric matrix $G$ with $+1$ upper off-diagonal and $-1$ lower off-diagonal entries, proving its eigenspace relates to a scaled quantum Fourier transform (QFT). The intuition for using $G$ is illustrated in Fig.~\ref{fig:Intuition}. Due to its exact QFT-based eigendecomposition, $G$ can admit efficient circuit implementations across quantum architectures.  

Furthermore, as shown in the figure, many antisymmetric matrices can be mapped to $G$ via signed permutations ($\pm 1$ elements). Thus, by parameterizing the eigendecomposition circuit for $e^G$ and adding a layer to approximate such permutations, one can approximate $e^A$ for any antisymmetric $A$.  
Since this framework requires no prior knowledge of $e^A$, it can also estimate $A$'s eigenspace and simulate $e^A$ directly.  

In the following sections, we first explain the eigendecomposition of the matrix $G$ and show how it can be mapped onto quantum circuits.

\section{Approximating Antisymmetric Matrices}
For a generic real antisymmetric matrix $A$, the elements satisfy $a_{ji} = -a_{ij}$ for $i \neq j$. The matrix formed by applying the sign function to $A$ (with zero diagonal) produces a matrix with equal numbers of $+1$ and $-1$ elements. For dense matrices, such matrices can be constructed by permuting rows/columns and flipping signs of elements in the following canonical form:
\begin{equation}
G = \begin{pmatrix}
    0 & 1 & 1 & \dots & 1 \\
    -1 & 0 & 1 & \dots & 1 \\
    -1 & -1 & 0 & \dots & 1 \\
    \vdots & \vdots & \vdots & \ddots & \vdots \\
    -1 & -1 & -1 & \dots & 0
\end{pmatrix}.
\end{equation}
This matrix represents a directed graph used to model tournaments without draws between competitors \cite{erdos1964problem,moon1968topics} and chiral quantum walks \cite{lu2016chiral}. In a transitive tournament, vertices admit a linear ordering (topological order) where every edge (off-diagonal entry) points from an earlier to a later vertex—yielding matrix $G$ with all $+1$ above and $-1$ below the diagonal. This sign pattern is acyclic (or "transitive" in tournament terminology).
When the sign patterns of a matrix contains cycles, we can render it acyclic by applying a signed permutation matrix ($\pm 1$ elements) to flip specific elements. Figure~\ref{fig:Intuition} illustrates this conversion for a random antisymmetric matrix to $G$'s canonical form using:
\begin{equation}
    P = \begin{pmatrix} 
    0 & 0 & 1 & 0 \\ 
    0 & -1 & 0 & 0 \\ 
    1 & 0 & 0 & 0 \\ 
    0 & 0 & 0 & 1 
    \end{pmatrix}.
\end{equation}
Thus, the sign matrix of $A$ can be expressed as $PGP^T$.

\subsection{Eigendecomposition of Uniform Antisymmetric Matrix $G$}

While antisymmetric matrices and their spectral properties are well-studied in linear algebra (e.g., diagonalizability with imaginary eigenvalues \cite{golub2013matrix}); to the best of our knowledge, the specific matrix \( G \)-defined by its uniform \(\pm 1\) transitive tournament structure-has not been explicitly analyzed in standard references. 
Structured antisymmetric matrices with patterned values similar to $G$, such as Toeplitz or circulant forms, often admit closed-form eigendecompositions involving trigonometric functions or roots of unity \cite{reichel1992eigenvalues,gray2006toeplitz}.  This occurs mostly because the roots of unity appear in the characteristic polynomial governing their eigenvalues.
In graph theory, antisymmetric matrices model tournaments, but spectral analyses typically emphasize graph invariants (e.g., skew energy \cite{adiga2010skew}) rather than explicit eigenpairs. 
Quantum literature involving the quantum Fourier transform frequently employs structured unitaries \cite{nielsen2010quantum}. However, to the best of our knowledge, no prior work represents the eigendecomposition of \( G \)  as a phase-shifted parameterized QFT.

\subsection{Eigendecomposition of Uniform Antisymmetric Matrix $G$}

The eigenspace of $G$ can be determined by solving the eigenvalue equation $G\vec{v} = \lambda \vec{v}$ explicitly in terms of vector components. For eigenvalue indices $k = 0, 1, \dots, N-1$, the eigenvalues and eigenvectors take the following form (see Appendix~\ref{appendix:Eigenvalues} for detailed derivation):

\begin{equation}
\lambda_k = -i \cot \left( \frac{(2k + 1)\pi}{2N} \right), \quad 
\vec{v}_k = \begin{bmatrix} 
1 \\ 
e^{i(2k + 1)\pi/N} \\ 
e^{i2(2k + 1)\pi/N} \\ 
\vdots \\ 
e^{i(N-1)(2k + 1)\pi/N} 
\end{bmatrix}.
\end{equation}

The normalized eigenvector components can be described as:
\begin{equation}
v_{k}^{(j)} = \frac{1}{\sqrt{N}} e^{i j \pi/N} \cdot \omega^{jk},
\end{equation}
where $\omega = e^{i2\pi/N}$. Defining $\beta = e^{i \pi / N}$, the normalized eigenvector matrix $V$ (with columns $\mathbf{v}_0, \mathbf{v}_1, \dots, \mathbf{v}_{N-1}$) relates to the quantum Fourier transform (QFT) as:
\begin{equation}
V = D \times F,
\end{equation}
where $D$ is a diagonal matrix with elements $d_{jj} = \beta^j$ and $F$ is the standard matrix for the quantum Fourier transform (see Appendix~\ref{appendix:QFTrelation} for proof). This structure forms a Vandermonde-like unitary matrix that differs from the standard QFT matrix by an additional phase shift $e^{ij\pi/N}$ in each column $j$.

The eigenvalues are purely imaginary for zero-diagonal $G$. When the diagonal contains elements $g$, the eigenvalues shift accordingly:
\begin{equation}
\lambda_k = g - i \cot \left( \frac{(2k + 1)\pi}{2N} \right).
\end{equation}

\subsection{Parameterized Quantum Circuit}

To approximate the exponential of an antisymmetric matrix $e^{A}$, we employ a parameterized circuit:

\begin{equation}
U(\theta) = P(\vec{\theta}_P)^\dagger D(\vec{\theta}_D)^\dagger F(\vec{\theta}_F)^\dagger \Lambda(\vec{\theta}_\Lambda) F(\vec{\theta}_F) D(\vec{\theta}_D) P(\vec{\theta}_P)
\end{equation}
where $U(\theta)$ represents the parameterized approximation of $e^A$. For notational simplicity, we use $\Lambda(\vec{\theta}_\Lambda) = e^\Lambda$ to denote the parameterized circuit implementing the exponential.
In terms of circuit, we can represent $U(\theta)$ as:
\[
\Qcircuit @C=0.5em @R=1em {
 &\qw / 
 & \multigate{1}{P(\vec{\theta}_P)} 
 & \multigate{1}{D(\vec{\theta}_D)} 
 & \multigate{1}{F(\vec{\theta}_F)}
 & \multigate{1}{\Lambda(\vec{\theta}_\Lambda)}
  & \multigate{1}{F(\vec{\theta}_F)^\dagger}
   & \multigate{1}{D(\vec{\theta}_D)^\dagger} 
    & \multigate{1}{P(\vec{\theta}_P)^\dagger} 
\\
  &\qw / 
   & \ghost{P(\vec{\theta}_P)} 
 & \ghost{D(\vec{\theta}_D)} 
 & \ghost{F(\vec{\theta}_F)}
 & \ghost{\Lambda(\vec{\theta}_\Lambda)}
  & \ghost{F(\vec{\theta}_F)^\dagger}
   & \ghost{D(\vec{\theta}_D)^\dagger} 
    & \ghost{P(\vec{\theta}_P)^\dagger} }
\]

The explicit circuit diagrams for each block component are provided below.

\subsubsection{Parameterized Circuits for Diagonal Matrices $D$ and $e^\Lambda$}

The diagonal matrix \( D \) has the form:
\begin{equation}
D = \text{diag}\left(e^{-i \frac{\pi}{2N}}, e^{-i \frac{3\pi}{2N}}, \dots, e^{-i \frac{(2N-1)\pi}{2N}}\right),
\end{equation}
where each entry represents a phase rotation \(\frac{(2j+1)\pi}{2N}\). For diagonal unitary matrices with structured phase relationships, we can employ a Kronecker product of $Z$-rotation gates as an approximation. The parameterized circuit for $D$ can be implemented as:
\begin{equation}
D(\vec{\theta}_D) = R_z(\phi_1) \otimes R_z(\phi_2) \otimes \dots \otimes R_z(\phi_n),
\end{equation}
where $R_z$ denotes the single-qubit rotation gate:
\begin{equation}
R_z(\phi) = \begin{pmatrix} 
e^{-i \phi/2} & 0 \\ 
0 & e^{i \phi/2} 
\end{pmatrix}
= e^{-i\phi/2}\begin{pmatrix}
1 & 0 \\ 
0 & e^{i \phi} 
\end{pmatrix}.
\end{equation}
Note that the second form is obtained by factoring out a global phase $e^{-i \phi/2}$, and all $\phi$ parameters represent real-valued angles.

Similarly we can also write the eigenvalue matrix \( \Lambda \) as a diagonal with entries  \( \Lambda \):
  \begin{equation}
  \Lambda = \text{diag}\left( -i \cot \left( \frac{\pi}{2N} \right), -i \cot \left( \frac{3\pi}{2N} \right), \dots, -i \cot \left( \frac{(2N-1)\pi}{2N} \right) \right)
\end{equation}
The matrix $\Lambda(\vec{\theta}_\Lambda) = e^\Lambda$ can be also approximated with a Kronecker tensor product of rotation \(Z\) gates on $n$ qubits:
\begin{equation}
\Lambda(\vec{\theta}_\Lambda) = R_z(\theta_1) \otimes R_z(\theta_2) \otimes \dots \otimes R_z(\theta_n)
\end{equation}
Note that a close approximation can be found to the original matrix \(\Lambda\) by using the diagonal entries  in the rotation gate angles. Since we will parameterize this circuit, we will not further analyze to find the exact angle values.

\subsubsection{Parameterized Circuit for $P$}
A rotation matrix $P$ that only flips the sign of elements and permutes the matrix can be implemented by using a sequence of controlled $X$ and $Z$ gates. To parameterize $P$,  we will consider it as a generic rotation with real elements and implement it by using a layered single and controlled rotation $Y$ gates.
\begin{figure}[ht]
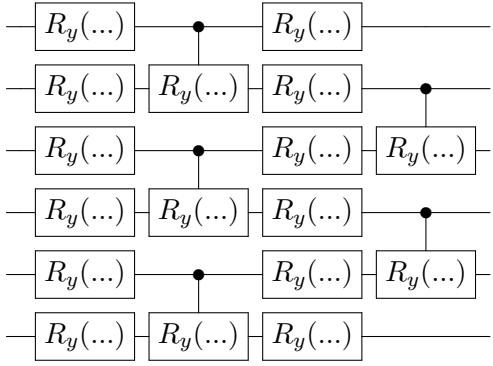
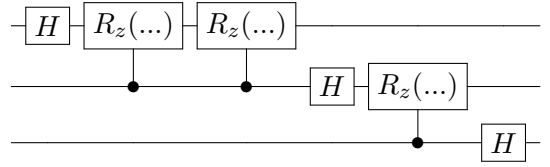

    \centering
    \begin{subfigure}{0.45\textwidth}
        \include{figcircuitP}
        \caption{Parameterized circuit implementation $P(\vec{\theta}_P)$.}
        \label{fig:circuitP}
    \end{subfigure}
    \hfill
    \begin{subfigure}{0.45\textwidth}
        \centering
        \include{figcircuitQFT}
        \caption{3-qubit parameterized QFT circuit $F(\vec{\theta}_F)$.}
        \label{fig:circuitQFT}
    \end{subfigure}
    \caption{Parameterized circuit implementations: (a) permutation matrix $P(\vec{\theta}_P)$ and (b) quantum Fourier transform $F(\vec{\theta}_F)$.}
    \label{fig:circuits}
\end{figure}

\subsubsection{Parameterized Quantum Fourier Transform $F$}
The quantum Fourier transform circuit $F$ can be parameterized by introducing adjustable rotation angles to the phase gates following each Hadamard gate in the standard QFT implementation.

\section{Numerical Simulation}
\subsection{Parameterized Circuit Layout}
In numerical simulations, the summary of our circuit setting is as follows:
\begin{enumerate}
    \item For the circuit $P$, we use only rotation $Y$ gates: one layer of single rotations, followed by a group of controlled rotations (entangling group of two qubits), then another layer of single gates, and finally another controlled rotations entangling previously  entangled group of twos.
    \item The circuit $D$ is designed as parameterized Kronecker product of rotation $Z$ gates.
    \item Then, we apply the standard quantum Fourier transform circuit where the rotation gates are parameterized.
    \item For the eigenvalues $\Lambda$, we use another Kronecker product of rotation $Z$ gates. Similarly to $D$, we use parameters as the angle values for the rotations, however, here, we take the cotangents of the parameters before feeding them to the rotation gates.
    \item We then apply the inverse circuits for each operator excluding $\Lambda$.
\end{enumerate}

\subsection{Optimization Criteria}
In the problem design we assume that an antisymmetric matrix $A$ is given. 
Therefore, after generating a circuit by using some $\vec{\theta}_k$ vector of angles, the logarithm of the circuit is compared with the matrix $A$ since in the assumption we do not know the unitary exponential $e^A$.
Note that the logarithm of the circuit $U(\vec{\theta}_k)$ can be approximated by taking the logarithm of the rotation $Z$s used for constructing the circuit $\Lambda$ because $PDF\Lambda F^\dagger D^\dagger P^\dagger$ is an eigendecomposition since the inverse of a circuit is the conjugate transpose, represented by the symbol ``$\dagger$", of a unitary matrix.

For an approximated matrix $\tilde{A}_k$ at the $k$th iteration of the optimization, we use the following loss function as an indication of error in the approximation:
\begin{equation}
    L(\vec{\theta}_k) = \|A-\tilde{A}_k\|_F =  \sqrt{\sum_i\sum_j |\Delta a_{ij} }|^2,
\end{equation}
where $\vec{\theta}_k$ optimization parameters used in the circuit $e^{\tilde{A}_k}$ and $\Delta a_{ij}$ are the error in the approximation of matrix element $a_{ij}$ of the target matrix $A$. In the simulations, the random instances are not normalized. Therefore, the loss can be very high. For instance, for a $\pm 1$ matrix, if the generated matrix has all $\pm 1$ elements but with opposite signs; $L(\vec{\theta}_k) = 2\sqrt{(N^2-N)}$, where $-N$ comes from the zero diagonal and $\pm2$ are the off-diagonal elements in the difference.

Here, note that if  the unitary matrix $U_A = e^A$ is available and we need to find the circuit approximation for the unitary, then we can similarly use $\|U_A -U_{\tilde{A}_k}\|_F$. However, to integrate global phases, using a fidelity based optimization criteria would be a better choice as done in many circuit decomposition techniques \cite{daskin2011decomposition}: For instance $1-|trace(U_A^\dagger U_{\tilde{A}_k})|/N$ or something similar.

\subsection{Test Cases and Simulation Results}
The simulation code is written by using Qiskit\cite{qiskit2024} Python library. An example run of the simulation code is as follows:
\begin{itemize}
    \item We first generate a matrix with randomly chosen (uniform distribution is used) matrix elements. For instance, the following matrix.
    \begin{equation}
        A = \begin{bmatrix}
 0.    &  0.4   &  0.3   & -0.35 \\
-0.4   &  0.    &  0.19  & -0.1  \\
-0.3   & -0.19  &  0.    & -0.27 \\
 0.35  &  0.1   &  0.27  &  0.   \\
\end{bmatrix}.
    \end{equation}

\item Then we run the circuit by initializing it with random parameters. For $\Lambda$ we have use the cotangent of the randomly generated phases as the initial parameters.
\item Then, in the optimization part, we use the minimization function in the SciPy\cite{2020SciPy-NMeth} library’s optimization tool with the method option ``L-BFGS-B" (one can use different methods. We have used this option because of limited cloud services.)
\item Then, the optimization is repeated until a determined maximum iteration number is reached or optimization converges.
\item The final exact circuit is drawn in Fig. which has the following unitary matrix form:
\begin{equation}
\label{eq:exampleUa}
   U_{A} =  \begin{bmatrix}
 0.82  &  0.33  &  0.26  & -0.38 \\
-0.42  &  0.9   &  0.11  & -0.04 \\
-0.28  & -0.25  &  0.9   & -0.2  \\
 0.27  &  0.13  &  0.32  &  0.9  \\
\end{bmatrix}.
\end{equation}
\begin{figure}
    \centering
    \includegraphics[width=\linewidth]{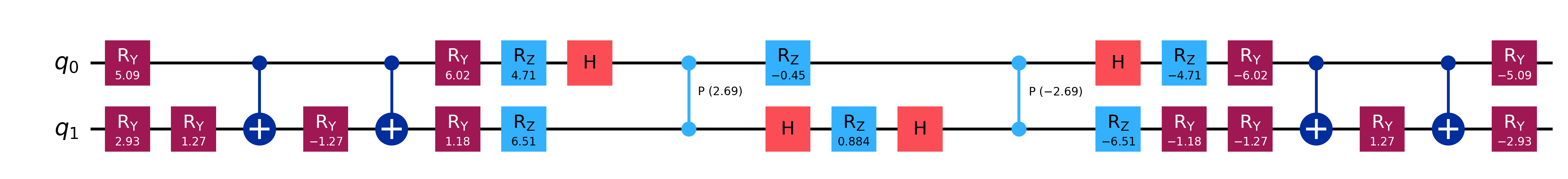}
    \caption{The example exact Qiskit circuit generated from optimization for the unitary $U_A = e^A$ given in Eq.\eqref{eq:exampleUa}. }
    \label{fig:enter-label}
\end{figure}
\end{itemize}

In order to see if this framework is able to approximate a matrix with a similar number of independent parameters,  we first use $4\times4$ matrices. 
The matrices are generated by setting the off-diagonal elements  on the upper part randomly from the range $[-1,1]$. While the off-diagonal elements in the lower part are set with the opposite signs to satisfy the antisymmetric property, the diagonal is kept zero.

To gauge the expressive power of the circuit better, we also employ random antisymmetric matrices with only $\pm 1$ off-diagonal elements with zero diagonal (dense case) and  matrices with $\{\pm 1, 0\}$ off-diagonal elements with zero diagonal (sparse case).
It is important to note that the matrices are not normalized; therefore $\|A\|_F$ can be as high as $N^2-N$ if all of its elements are $\pm 1$.

\begin{figure}[ht]
    \centering
    \begin{subfigure}{0.4\textwidth}
        \includegraphics[width=1\linewidth]{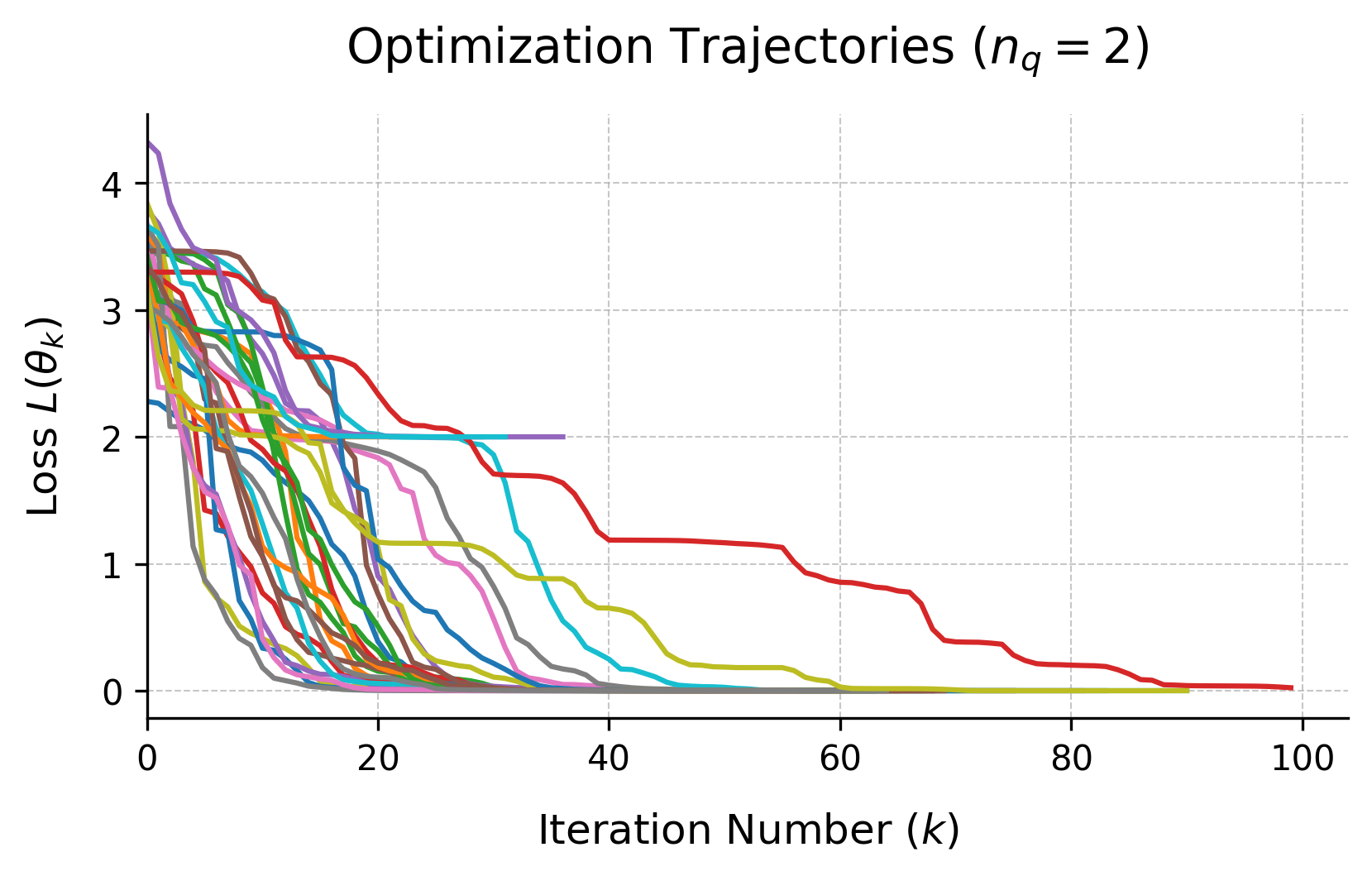}
        \caption{Matrices with off diagonal elements from \{-1,1\}.}
    \end{subfigure}~
    \begin{subfigure}{0.4\textwidth}
        \includegraphics[width=1\linewidth]{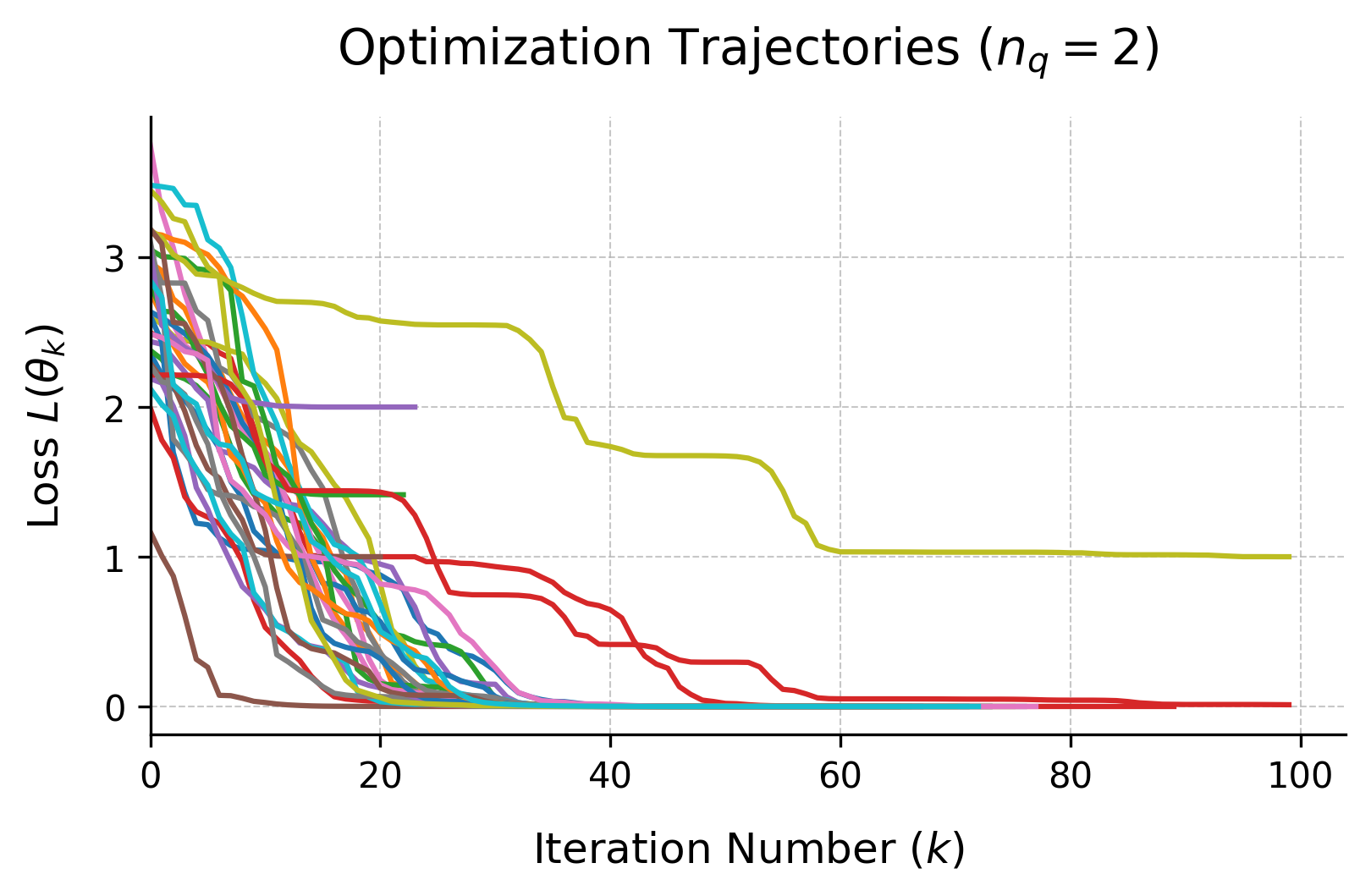}
        \caption{Matrices with off diagonal elements from \{-1,1,0\}.}
    \end{subfigure}\\
    \begin{subfigure}{0.4\textwidth}
        \includegraphics[width=1\linewidth]{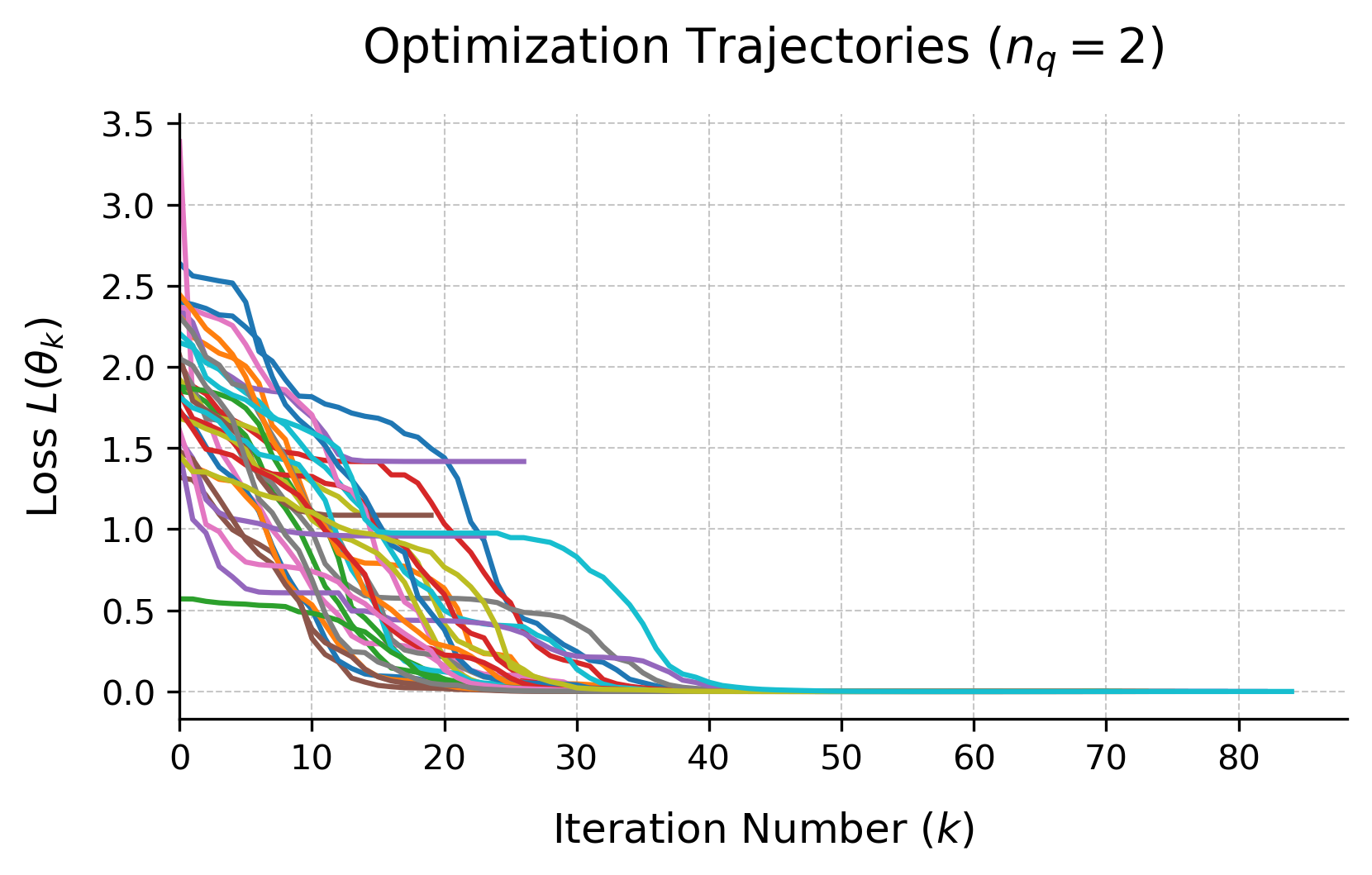}
        \caption{Matrices with off diagonal real elements in range [-1,1].}
    \end{subfigure}
    \begin{subfigure}{0.4\textwidth}
        \includegraphics[width=1\linewidth]{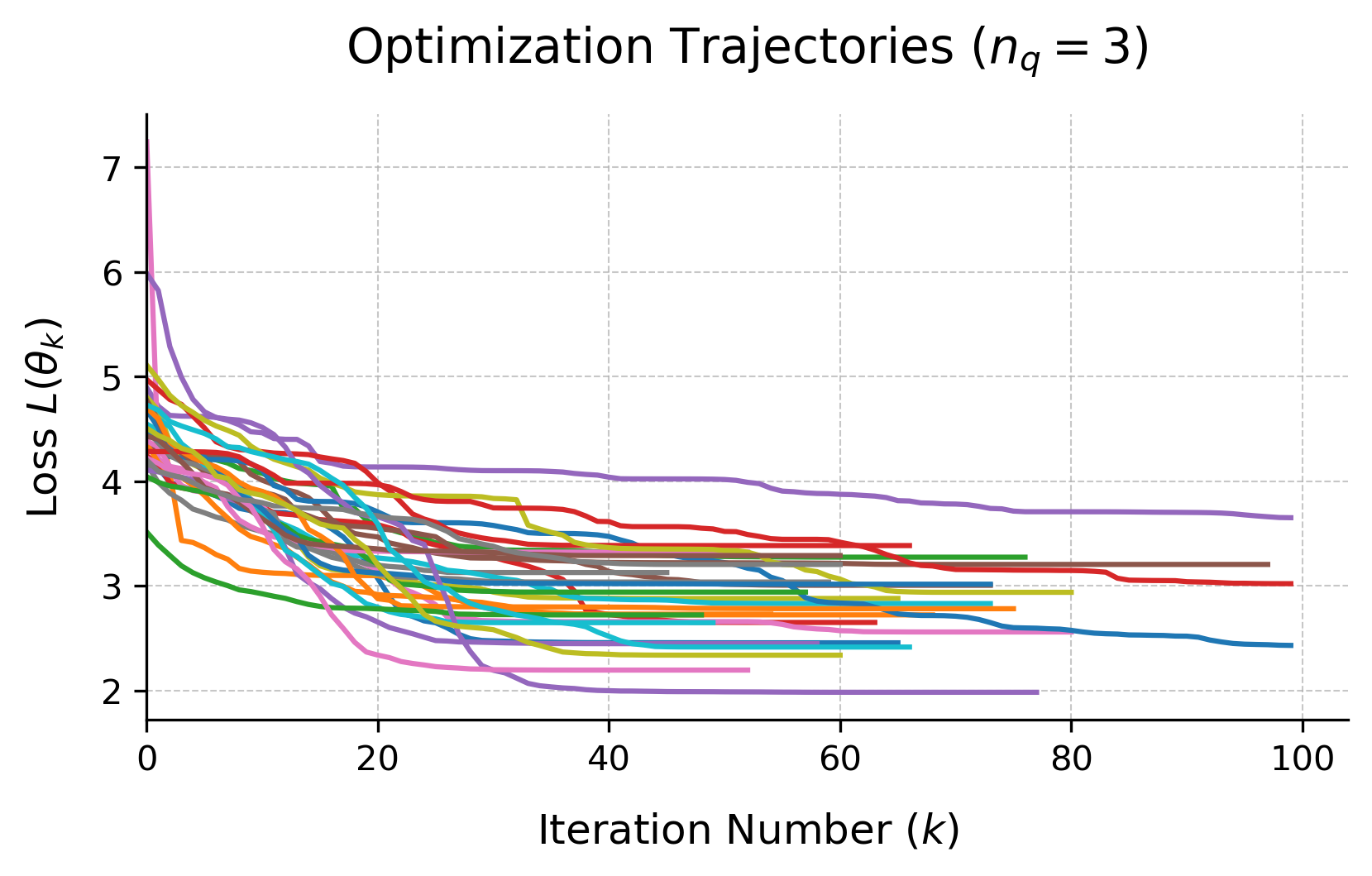}
        \caption{Matrices with off diagonal real elements in range [-1,1].}
    \end{subfigure}
    \caption{The loss during optimization for 30 random  matrices of dimension $2^{n_q}$, where $n_q=2 \text{ and } 3$ is the number of qubits required for the matrix. $L(\vec{\theta}_k) = \|A-\tilde{A}_k\|_F$: The matrices are not normalized in the loss.}
    \label{fig:simresults2by2}
\end{figure}

Fig.\ref{fig:simresults2by2} shows the results for random matrices for these cases. As it can be seen from the figure, the exact circuit is found in almost all cases but a few. The few unsuccessful runs are mostly related to optimization tools being stuck for the local minima which can be prevented in some cases by restarting optimization by using different initial parameters.

\begin{figure}[ht]
    \centering
    \begin{subfigure}{0.4\textwidth}
    \includegraphics[width=1\linewidth]{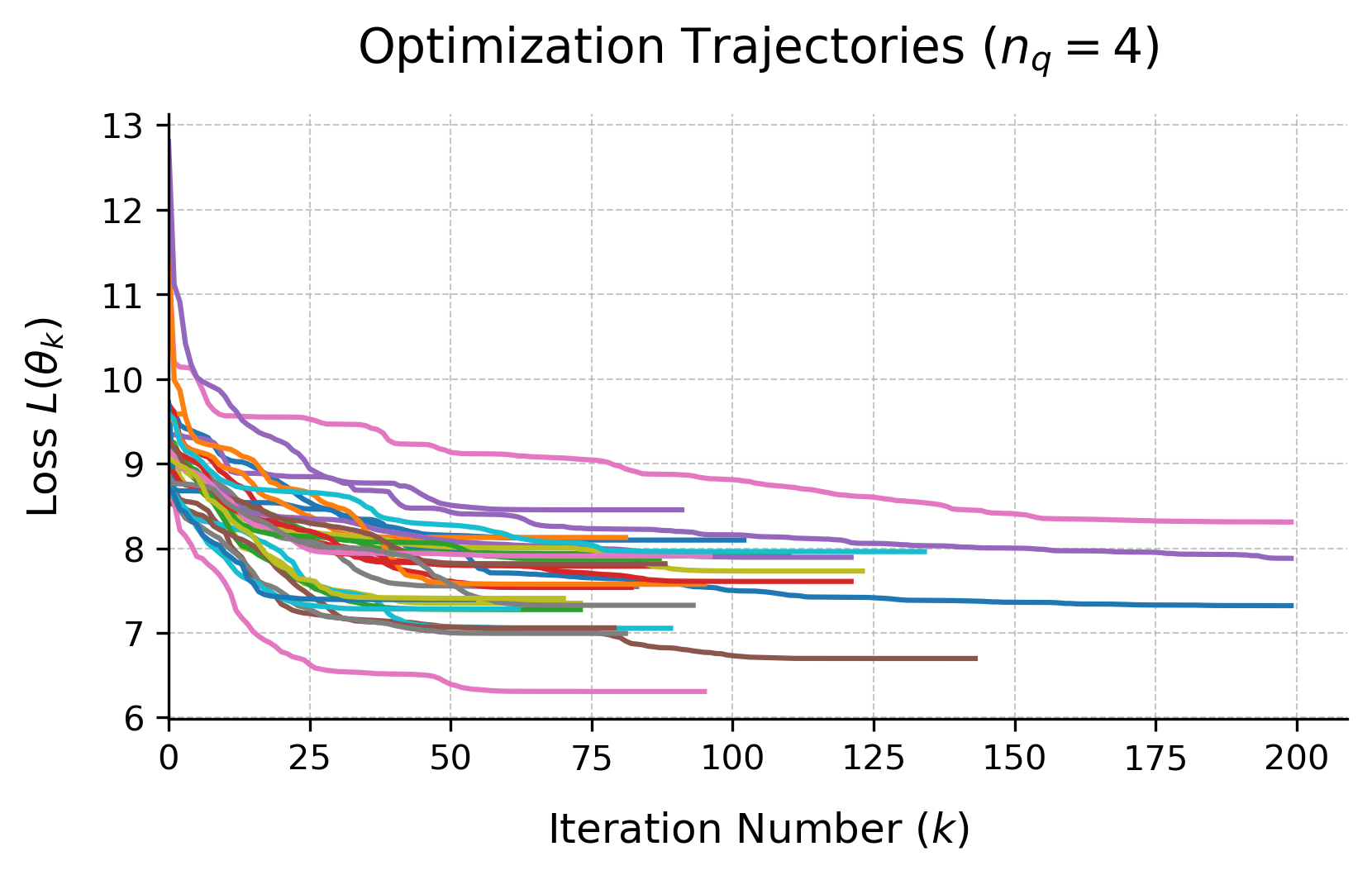}
    \caption{Matrices with off diagonal real elements in range [-1,1].}
    \end{subfigure}~
    \begin{subfigure}{0.4\textwidth}
        \includegraphics[width=1\linewidth]{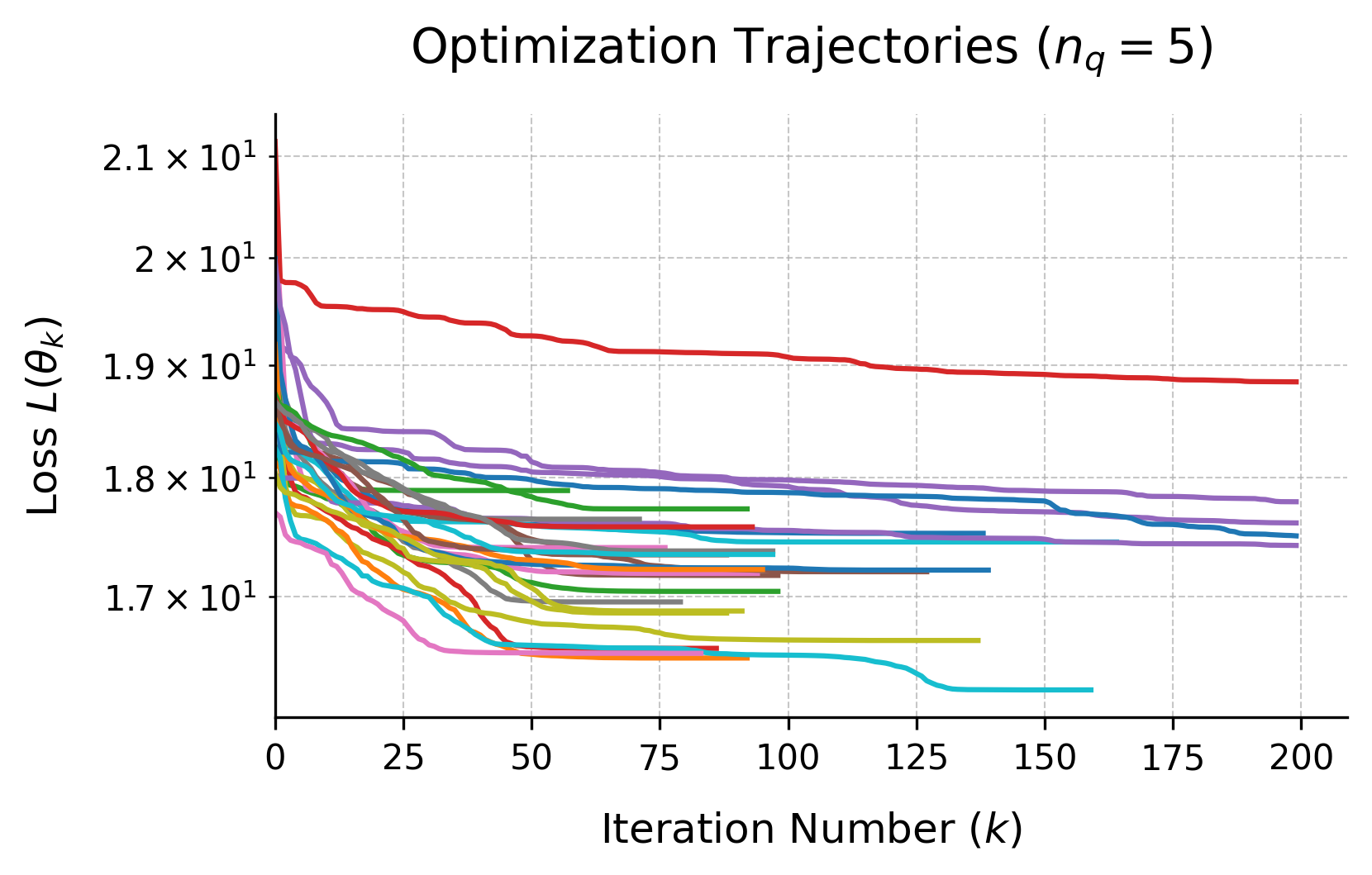}
       \caption{Matrices with off diagonal real elements in range [-1,1].}
    \end{subfigure}\\
    \begin{subfigure}{0.4\textwidth}
        \includegraphics[width=1\linewidth]{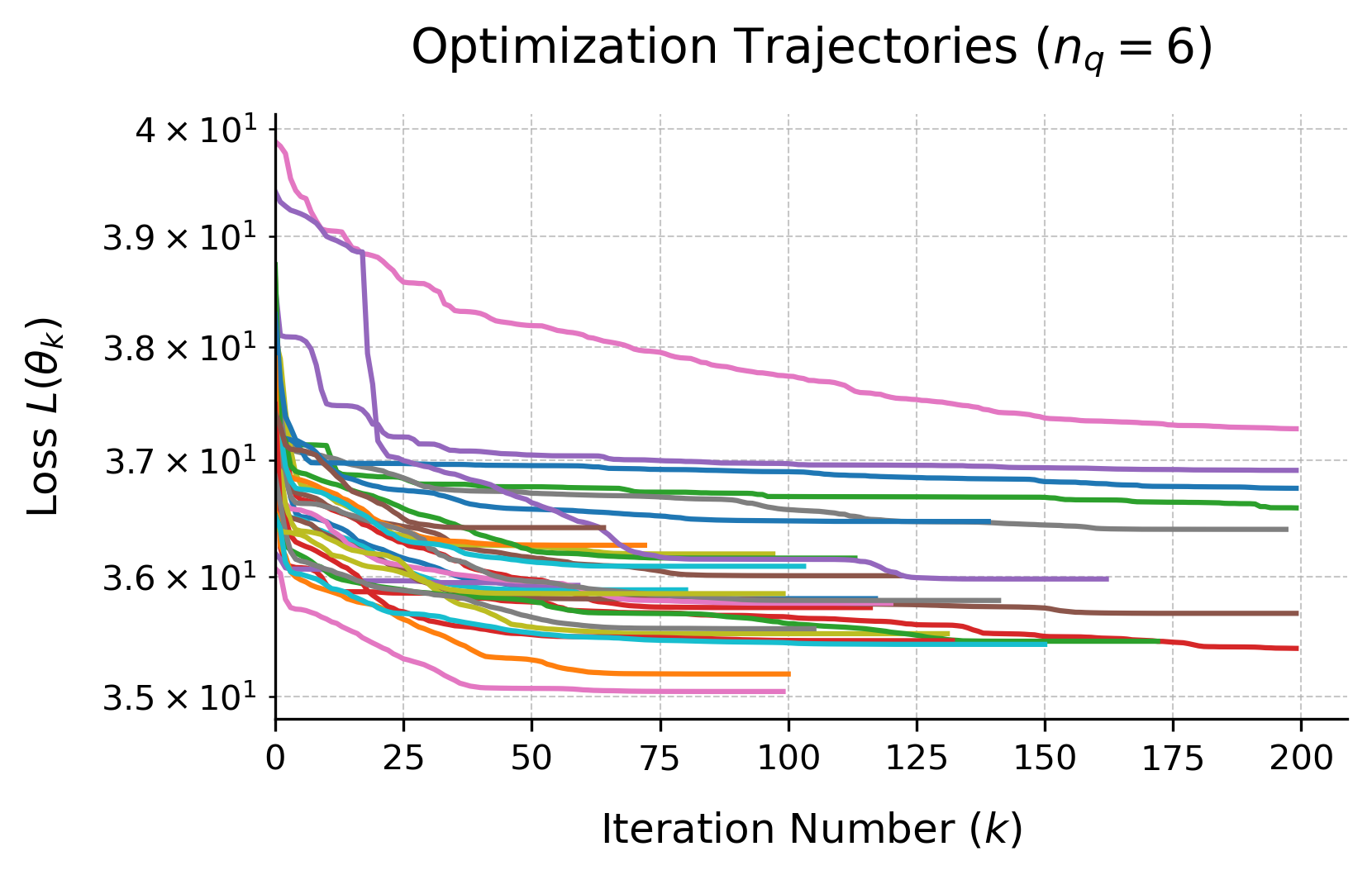}
 \caption{Matrices with off diagonal real elements in range [-1,1].}
    \end{subfigure}~
        \begin{subfigure}{0.4\textwidth}
        \includegraphics[width=1\linewidth]{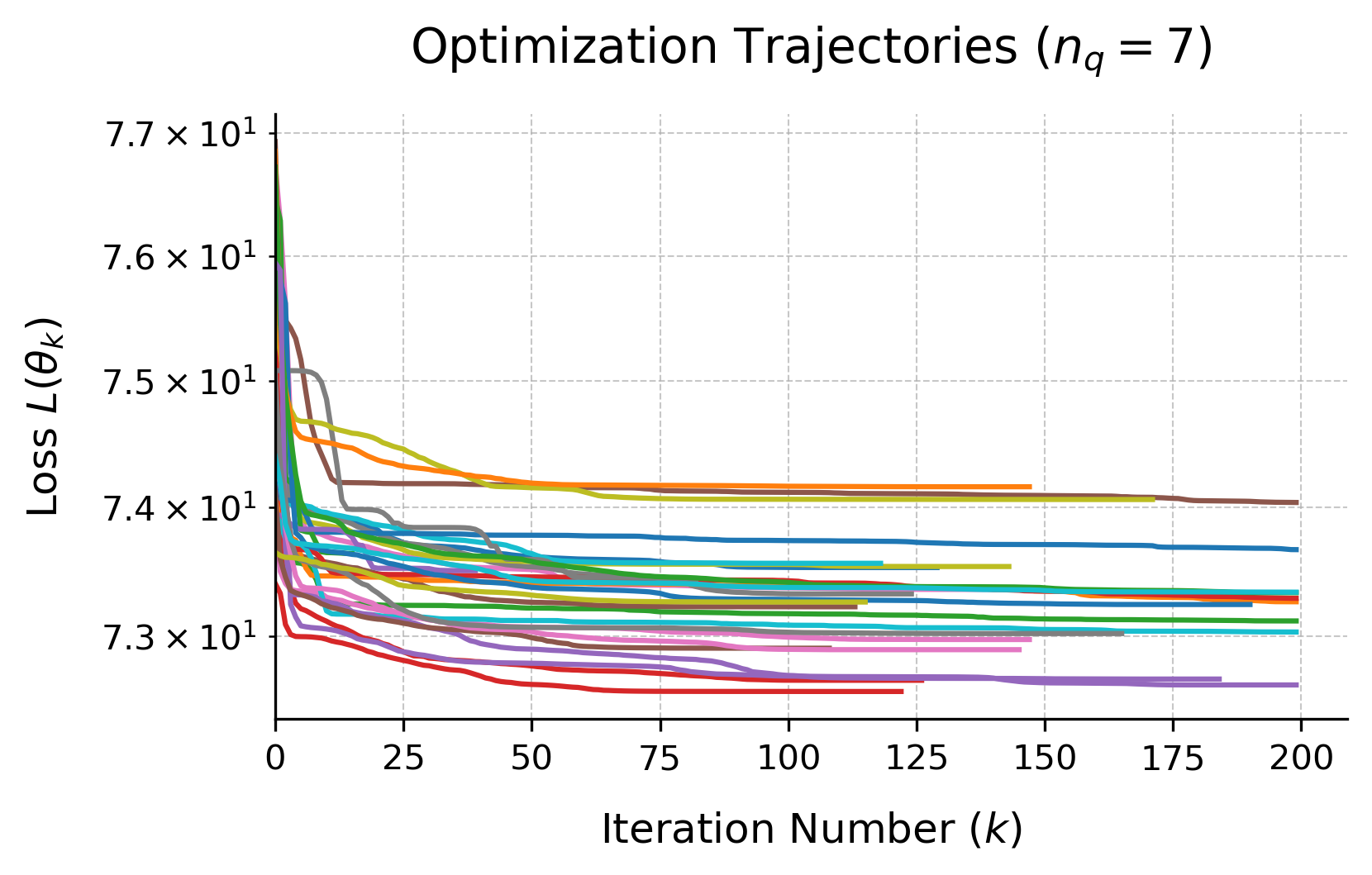}
        \caption{Matrices with off diagonal real elements in range [-1,1].}
    \end{subfigure}
    \caption{ For each $n_q = 4, \dots, 7$; the loss during optimization for 30 random matrices of dimension $2^{n_q}$. $L(\vec{\theta}_k) = \|A-\tilde{A}_k\|_F$: The matrices are not normalized in the loss.   }
    \label{fig:simresultsnbyn}
\end{figure}

We also show the similar simulation results for larger matrices with random elements from range 
$[-1,1]$.  We use the same Frobenius norm for the loss function. As seen in Fig.\ref{fig:simresultsnbyn}, the loss function is larger with the matrix sizes. This can be related to the unnormalized loss function and also the expressive power of the circuit. Since for larger dimensions, the gap between $n$ and $2^n$ gets larger. Therefore, the circuit is no longer able to express the full eigendecomposition of the matrix. 

\section{Future Directions and Conclusion}

 Since antisymmetric matrices are diagonalizable with purely imaginary eigenvalues, it is well-known that classical methods like the Schur decomposition or orthogonal transformations (e.g., Householder reflections) can simplify exponentiation \cite{golub2013matrix}. The scaling-and-squaring algorithm, combined with Padé approximants \cite{higham2005scaling} for dense matrices and Krylov subspace methods for large sparse systems \cite{moler2003nineteen} can be used to find the exponential of the matrix. 
 However, these methods scale polynomially with matrix size which is exponential in the number of qubits. Therefore, this  limits their utility for high-dimensional quantum simulations.  

 In this paper we present a generic framework with $O(n^2)$ quantum gates which can be used to design quantum circuits for a known unitary $U$ or it can be used to find $U=e^A$ and so the circuit for the simulation of an antisymmetric matrix $A$.
 In addition, it provides an algorithmic framework to estimate the eigendecomposition of a matrix $A$. If the matrix $A$ is also unknown, it can be employed to emulate the behavior of the unknown antisymmetric quantum system $A$.
 
However, using this framework without modification for larger matrices may require additional settings: The quantum circuit introduced in this paper comprises $O(n^2)$ gates without repetitions. A generic $n$-qubit unitary matrix typically requires $\sim 2^{2n}$ elementary quantum gates for decomposition without ancilla qubits (see, e.g., lower-bound analyses in \cite{shende2005synthesis}). This exponential scaling in the number of qubits  can be proven by using a simple parameter counting argument \cite{nielsen2010quantum}. As a result of this exponential complexity requirement, a full rank approximation of a generic matrix would necessitate exponentially many gates.  

As a result, the framework can be used to partially (such as only the principal components) approximate eigendecomposition of matrices. However, in approximating larger matrices, it is mathematically difficult (and in some cases impossible) to achieve arbitrary precision below a certain error threshold without adding more parameterized gates to the circuit structure.  
For example, a 3-qubit circuit has approximately parameterized 17 gates that try to predict a $64$-element unitary matrix.  This limitation can be addressed potentially by seeking lower-rank approximations of the target matrix or expanding the expressivity of the circuit by adding gates to existing layers, introducing new parameterized layers.

\section{Data availability}
The simulation code used to generate all figures presented in this paper  is publicly available
on: \url{https://github.com/adaskin/pqc-antisymmetric-matrix.git}
\section{Funding}
This project is not funded by any funding agency.

\appendix{

\section{Eigenvalues of $G$}
\label{appendix:Eigenvalues}
For an \( n \times n \) matrix \( G \), let us denote the entries as:
\begin{equation}
G_{ij} = 
\begin{cases} 
g_i & \text{if } i = j, \\
+1 & \text{if } i < j, \\
-1 & \text{if } i > j.
\end{cases}
\end{equation}

For \( \mathbf{v} = [v_1, v_2, \dots, v_n]^T \), from the eigenvalue equation \( G \mathbf{v} = \lambda \mathbf{v} \), we can write the followings for each row:
\begin{align}
&\text{For \( i = 1, \) }   & g_1 v_1 + v_2 + v_3 + \dots + v_n = \lambda v_1.\\
&\text{For \( i = 2, \) }   & -v_1 + g_2 v_2 + v_3 + \dots + v_n = \lambda v_2.\\
&\text{For general  \( i, \) }   & -v_1 - v_2 - \dots - v_{i-1} + g_i v_i + v_{i+1} + \dots + v_n = \lambda v_i.\\
&\text{For \( i = n, \) }    &-v_1 - v_2 - \dots - v_{n-1} + g_n v_n = \lambda v_n.
\end{align}

If all diagonal elements are equal (\( g_1 = g_2 = \dots = d_n = g \)), the matrix has a special structure where the eigenvalues and eigenvectors can be derived  easily from the above equations.
For instance assume that $g=0$, i.e. all diagonal elements are \( 0 \). The eigenvalue equation gives the followings:
:
\begin{align}
&\text{For \( i = 1, \) }   & v_2 + v_3 + \dots + v_n = \lambda v_1.\\
&\text{For \( i = 2, \) }   & -v_1 + v_3 + \dots + v_n = \lambda v_2.\\
&\text{For general  \( i, \) }   & -v_1 - v_2 - \dots - v_{i-1} + v_{i+1} + \dots + v_n = \lambda v_i.\\
&\text{For \( i = n, \) }    & -v_1 - v_2 - \dots - v_{n-1} = \lambda v_n.
\end{align}

If we subtract the \( i \)-th equation from the \( (i+1) \)-th equation:
\begin{equation}
- v_i - v_{i+1} = \lambda (v_{i+1} - v_i) = \lambda v_{i+1} - \lambda v_i.
\end{equation}
This simplifies as:
\begin{equation}
(\lambda - 1) v_i = (\lambda + 1) v_{i+1}.
\end{equation}
Thus, we can write the following ratio:
\begin{equation}
\frac{v_{i+1}}{v_i} = \frac{\lambda - 1}{\lambda + 1}
\end{equation}
Since this ratio is constant for all \( i \),  the eigenvector \( \mathbf{v} \) has the following form:
\begin{equation}
v_i = \left( \frac{\lambda - 1}{\lambda + 1} \right)^{i-1} v_1.
\end{equation}
Now, substitute this into the last equation (\( i = n \)):
\begin{equation}
- v_1 \left( 1 + \frac{\lambda - 1}{\lambda + 1} + \left( \frac{\lambda - 1}{\lambda + 1} \right)^2 + \dots + \left( \frac{\lambda - 1}{\lambda + 1} \right)^{n-2} \right) = \lambda v_n.
\end{equation}
Since \( v_n = \left( \frac{\lambda - 1}{\lambda + 1} \right)^{n-1} v_1 \), we have:
\begin{equation}
- v_1 \left( \frac{1 - \left( \frac{\lambda - 1}{\lambda + 1} \right)^{n-1}}{1 - \frac{\lambda - 1}{\lambda + 1}} \right) = \lambda \left( \frac{\lambda - 1}{\lambda + 1} \right)^{n-1} v_1.
\end{equation}
Simplifying the denominator:
\begin{equation}
1 - \frac{\lambda - 1}{\lambda + 1} = \frac{2}{\lambda + 1}.
\end{equation}
Thus:
\begin{equation}
- v_1 \left( \frac{(\lambda + 1)}{2} \left( 1 - \left( \frac{\lambda - 1}{\lambda + 1} \right)^{n-1} \right) \right) = \lambda \left( \frac{\lambda - 1}{\lambda + 1} \right)^{n-1} v_1.
\end{equation}
Cancel \( v_1 \) (assuming \( v_1 \neq 0 \)):
\begin{equation}
- \frac{\lambda + 1}{2} + \frac{\lambda + 1}{2} \left( \frac{\lambda - 1}{\lambda + 1} \right)^{n-1} = \lambda \left( \frac{\lambda - 1}{\lambda + 1} \right)^{n-1}.
\end{equation}
Multiply through by 2:
\begin{equation}
- (\lambda + 1) + (\lambda + 1) \left( \frac{\lambda - 1}{\lambda + 1} \right)^{n-1} = 2 \lambda \left( \frac{\lambda - 1}{\lambda + 1} \right)^{n-1}.
\end{equation}
Combine terms:
\begin{equation}
- (\lambda + 1) = \left( 2 \lambda - (\lambda + 1) \right) \left( \frac{\lambda - 1}{\lambda + 1} \right)^{n-1}.
\end{equation}
\begin{equation}
- (\lambda + 1) = (\lambda - 1) \left( \frac{\lambda - 1}{\lambda + 1} \right)^{n-1}.
\end{equation}
\begin{equation}
- (\lambda + 1) = (\lambda - 1)^n / (\lambda + 1)^{n-1}.
\end{equation}
This gives us the following final equation:
\begin{equation}
(\lambda + 1)^n + (\lambda - 1)^n = 0.
\end{equation}
This can be rewritten as:
\begin{equation}
\left( \frac{\lambda + 1}{\lambda - 1} \right)^n = -1.
\end{equation}
Let \( z = \frac{\lambda + 1}{\lambda - 1} \), then \( z^n = -1 \), so \( z = e^{i (2k + 1) \pi / n} \) for \( k = 0, 1, \dots, n-1 \). Solving for \( \lambda \):
\begin{equation}
z (\lambda - 1) = \lambda + 1.
\end{equation}
\begin{equation}
\lambda (z - 1) = z + 1.
\end{equation}
\begin{equation}
\lambda = \frac{z + 1}{z - 1}.
\end{equation}
Substituting \( z = e^{i (2k + 1) \pi / n} \):
\begin{equation}
\lambda_k = \frac{e^{i (2k + 1) \pi / n} + 1}{e^{i (2k + 1) \pi / n} - 1}.
\end{equation}
Multiply numerator and denominator by \( e^{-i (2k + 1) \pi / (2n)} \):
\begin{equation}
\lambda_k = \frac{e^{i (2k + 1) \pi / (2n)} + e^{-i (2k + 1) \pi / (2n)}}{e^{i (2k + 1) \pi / (2n)} - e^{-i (2k + 1) \pi / (2n)}}.
\end{equation}
Using \( \cos \theta = \frac{e^{i \theta} + e^{-i \theta}}{2} \) and \( \sin \theta = \frac{e^{i \theta} - e^{-i \theta}}{2i} \):
\begin{equation}
\lambda_k = \frac{2 \cos \left( \frac{(2k + 1) \pi}{2n} \right)}{2 i \sin \left( \frac{(2k + 1) \pi}{2n} \right)} = -i \cot \left( \frac{(2k + 1) \pi}{2n} \right), 
\end{equation}
From the initial equations, we know that the eigenvector corresponding to \( \lambda_k \) has the following components:
\begin{equation}
v_i = \left( \frac{\lambda_k - 1}{\lambda_k + 1} \right)^{i-1} v_1.
\end{equation}
Substituting \( \lambda_k = -i \cot \theta_k \) where \( \theta_k = \frac{(2k + 1) \pi}{2n} \):
\begin{equation}
\frac{\lambda_k - 1}{\lambda_k + 1} = \frac{-i \cot \theta_k - 1}{-i \cot \theta_k + 1} = \frac{-i \cos \theta_k / \sin \theta_k - 1}{-i \cos \theta_k / \sin \theta_k + 1} = \frac{-i \cos \theta_k - \sin \theta_k}{-i \cos \theta_k + \sin \theta_k}.
\end{equation}
Multiplying the numerator and the denominator by \( i \) gives us the following:
\begin{equation}
= \frac{ -i (\sin \theta_k + i \cos \theta_k ) }{ i (\sin \theta_k - i \cos \theta_k ) } = \frac{ -i \sin \theta_k + \cos \theta_k }{ i \sin \theta_k + \cos \theta_k } = \frac{ e^{i \theta_k} }{ e^{-i \theta_k} } = e^{i 2 \theta_k}.
\end{equation}
Therefore, we can write:
\begin{equation}
\frac{\lambda_k - 1}{\lambda_k + 1} = e^{i 2 \theta_k} = e^{i (2k + 1) \pi / n}.
\end{equation}
So the eigenvector components are:
\begin{equation}
v_i = e^{i (2k + 1) \pi (i - 1) / n} v_1.
\end{equation}
Setting \( v_1 = 1 \), the eigenvector  can be written as:
\begin{equation}
\mathbf{v}_k = \begin{bmatrix} 1 \\ e^{i (2k + 1) \pi / n} \\ e^{i 2 (2k + 1) \pi / n} \\ \vdots \\ e^{i (n-1) (2k + 1) \pi / n} \end{bmatrix}.
\end{equation}
If the diagonal elements are not zero but all \( g_i = g\), the eigenvalues are simply shifted by \( g \):
\begin{equation}
\lambda_k =g - i \cot \left( \frac{(2k + 1) \pi}{2n} \right).
\end{equation}

\section{Relating $V$ to the Quantum Fourier Transform}
\label{appendix:QFTrelation}
For \( \beta = e^{i \pi / N} \) and \( k = 0, 1, \dots, n-1 \), we can write the normalized matrix $V$ with columns \( \mathbf{v}_0, \mathbf{v}_1, \dots, \mathbf{v}_{n-1} \) as:
\begin{equation}
V = \frac{1}{\sqrt{n}} \begin{bmatrix}
\beta^{0(2 \cdot 0 + 1)} & \beta^{0(2 \cdot 1 + 1)} & \cdots & \beta^{0(2(n-1) + 1)} \\
\beta^{1(2 \cdot 0 + 1)} & \beta^{1(2 \cdot 1 + 1)} & \cdots & \beta^{1(2(n-1) + 1)} \\
\vdots & \vdots & \ddots & \vdots \\
\beta^{(n-1)(2 \cdot 0 + 1)} & \beta^{(n-1)(2 \cdot 1 + 1)} & \cdots & \beta^{(n-1)(2(n-1) + 1)}
\end{bmatrix}.
\end{equation}
Using $\omega = e^{i2\pi/N}$, we can write this matrix explicitly to relate with quantum Fourier transform.
\begin{align}
&V = \frac{1}{\sqrt{N}} \begin{bmatrix}
\beta^0\omega^0 & \beta^0\omega^0 &\beta^0\omega^0 & \cdots &\beta^0\omega^0 \\
\beta^1\omega^0 & \beta^1\omega^1 &\beta^1\omega^2& \cdots & \beta^1\omega^{(n-1)} \\
\vdots & \vdots & \vdots & \ddots & \vdots \\
\beta^{(n-1)}\omega^{(n-1)} & \beta^{(n-1)}\omega^{1(n-1)} & \beta^{(n-1)}\omega^{2(n-1)}& \cdots & \beta^{(n-1)}\omega^{(n-1)(n-1)}
\end{bmatrix}.
\end{align}
Therefore, $V= D\times F$ is the scaled version of the quantum Fourier transform matrix $F$.

}

\bibliographystyle{unsrtnat}
\bibliography{ref}

\begin{thebibliography}{41}
\providecommand{\natexlab}[1]{#1}
\providecommand{\url}[1]{\texttt{#1}}
\expandafter\ifx\csname urlstyle\endcsname\relax
  \providecommand{\doi}[1]{doi: #1}\else
  \providecommand{\doi}{doi: \begingroup \urlstyle{rm}\Url}\fi

\bibitem[Lloyd(1996)]{lloyd1996universal}
Seth Lloyd.
\newblock Universal quantum simulators.
\newblock \emph{Science}, 273\penalty0 (5278):\penalty0 1073--1078, 1996.

\bibitem[Nielsen and Chuang(2010)]{nielsen2010quantum}
Michael~A Nielsen and Isaac~L Chuang.
\newblock \emph{Quantum computation and quantum information}.
\newblock Cambridge university press, 2010.

\bibitem[Daskin et~al.(2012)Daskin, Grama, Kollias, and Kais]{daskin2012universal}
Anmer Daskin, Ananth Grama, Giorgos Kollias, and Sabre Kais.
\newblock Universal programmable quantum circuit schemes to emulate an operator.
\newblock \emph{The Journal of chemical physics}, 137\penalty0 (23), 2012.

\bibitem[Low and Chuang(2019)]{low2019hamiltonian}
Guang~Hao Low and Isaac~L Chuang.
\newblock Hamiltonian simulation by qubitization.
\newblock \emph{Quantum}, 3:\penalty0 163, 2019.

\bibitem[Gily{\'e}n et~al.(2019)Gily{\'e}n, Su, Low, and Wiebe]{gilyen2019quantum}
Andr{\'a}s Gily{\'e}n, Yuan Su, Guang~Hao Low, and Nathan Wiebe.
\newblock Quantum singular value transformation and beyond: exponential improvements for quantum matrix arithmetics.
\newblock In \emph{Proceedings of the 51st annual ACM SIGACT symposium on theory of computing}, pages 193--204, 2019.

\bibitem[Peruzzo et~al.(2014)Peruzzo, McClean, Shadbolt, Yung, Zhou, Love, Aspuru-Guzik, and O’brien]{peruzzo2014variational}
Alberto Peruzzo, Jarrod McClean, Peter Shadbolt, Man-Hong Yung, Xiao-Qi Zhou, Peter~J Love, Al{\'a}n Aspuru-Guzik, and Jeremy~L O’brien.
\newblock A variational eigenvalue solver on a photonic quantum processor.
\newblock \emph{Nature communications}, 5\penalty0 (1):\penalty0 4213, 2014.

\bibitem[Romero et~al.(2017)Romero, Olson, and Aspuru-Guzik]{romero2017quantum}
Jonathan Romero, Jonathan~P Olson, and Alan Aspuru-Guzik.
\newblock Quantum autoencoders for efficient compression of quantum data.
\newblock \emph{Quantum Science and Technology}, 2\penalty0 (4):\penalty0 045001, 2017.

\bibitem[Bauer et~al.(2016)Bauer, Wecker, Millis, Hastings, and Troyer]{bauer2016hybrid}
Bela Bauer, Dave Wecker, Andrew~J Millis, Matthew~B Hastings, and Matthias Troyer.
\newblock Hybrid quantum-classical approach to correlated materials.
\newblock \emph{Physical Review X}, 6\penalty0 (3):\penalty0 031045, 2016.

\bibitem[Tilly et~al.(2022)Tilly, Chen, Cao, Picozzi, Setia, Li, Grant, Wossnig, Rungger, Booth, et~al.]{tilly2022variational}
Jules Tilly, Hongxiang Chen, Shuxiang Cao, Dario Picozzi, Kanav Setia, Ying Li, Edward Grant, Leonard Wossnig, Ivan Rungger, George~H Booth, et~al.
\newblock The variational quantum eigensolver: a review of methods and best practices.
\newblock \emph{Physics Reports}, 986:\penalty0 1--128, 2022.

\bibitem[Benedetti et~al.(2019)Benedetti, Lloyd, Sack, and Fiorentini]{benedetti2019parameterized}
Marcello Benedetti, Erika Lloyd, Stefan Sack, and Mattia Fiorentini.
\newblock Parameterized quantum circuits as machine learning models.
\newblock \emph{Quantum science and technology}, 4\penalty0 (4):\penalty0 043001, 2019.

\bibitem[Cerezo et~al.(2022)Cerezo, Verdon, Huang, Cincio, and Coles]{cerezo2022challenges}
Marco Cerezo, Guillaume Verdon, Hsin-Yuan Huang, Lukasz Cincio, and Patrick~J Coles.
\newblock Challenges and opportunities in quantum machine learning.
\newblock \emph{Nature computational science}, 2\penalty0 (9):\penalty0 567--576, 2022.

\bibitem[Combarro et~al.(2023)Combarro, Gonz{\'a}lez-Castillo, and Di~Meglio]{combarro2023practical}
Elias~F Combarro, Samuel Gonz{\'a}lez-Castillo, and Alberto Di~Meglio.
\newblock \emph{A practical guide to quantum machine learning and quantum optimization: Hands-on approach to modern quantum algorithms}.
\newblock Packt Publishing Ltd, 2023.

\bibitem[Adiga et~al.(2010)Adiga, Balakrishnan, and So]{adiga2010skew}
Chandrashekar Adiga, R~Balakrishnan, and Wasin So.
\newblock The skew energy of a digraph.
\newblock \emph{Linear Algebra and its Applications}, 432\penalty0 (7):\penalty0 1825--1835, 2010.

\bibitem[Hayashi et~al.(2022)Hayashi, Aksoy, and Park]{hayashi2022skew}
Koby Hayashi, Sinan~G Aksoy, and Haesun Park.
\newblock Skew-symmetric adjacency matrices for clustering directed graphs.
\newblock In \emph{2022 IEEE International Conference on Big Data (Big Data)}, pages 555--564. IEEE, 2022.

\bibitem[Cavers et~al.(2012)Cavers, Cioab{\u{a}}, Fallat, Gregory, Haemers, Kirkland, McDonald, and Tsatsomeros]{cavers2012skew}
Michael Cavers, SM~Cioab{\u{a}}, Shaun Fallat, DA~Gregory, WH~Haemers, SJ~Kirkland, JJ~McDonald, and Michael Tsatsomeros.
\newblock Skew-adjacency matrices of graphs.
\newblock \emph{Linear algebra and its applications}, 436\penalty0 (12):\penalty0 4512--4529, 2012.

\bibitem[Li and Lian(2013)]{li2013survey}
Xueliang Li and Huishu Lian.
\newblock A survey on the skew energy of oriented graphs.
\newblock \emph{arXiv preprint arXiv:1304.5707}, 2013.

\bibitem[Kovac and Kressner(2017)]{kovac2017structure}
Erna~Begovic Kovac and Daniel Kressner.
\newblock Structure-preserving low multilinear rank approximation of antisymmetric tensors.
\newblock \emph{SIAM journal on matrix analysis and applications}, 38\penalty0 (3):\penalty0 967--983, 2017.

\bibitem[Ceruti and Lubich(2020)]{ceruti2020time}
Gianluca Ceruti and Christian Lubich.
\newblock Time integration of symmetric and anti-symmetric low-rank matrices and tucker tensors.
\newblock \emph{BIT Numerical Mathematics}, 60:\penalty0 591--614, 2020.

\bibitem[Klus et~al.(2021)Klus, Gel{\ss}, N{\"u}ske, and No{\'e}]{klus2021symmetric}
Stefan Klus, Patrick Gel{\ss}, Feliks N{\"u}ske, and Frank No{\'e}.
\newblock Symmetric and antisymmetric kernels for machine learning problems in quantum physics and chemistry.
\newblock \emph{Machine Learning: Science and Technology}, 2\penalty0 (4):\penalty0 045016, 2021.

\bibitem[Trabelsi et~al.(2017)Trabelsi, Bilaniuk, Zhang, Serdyuk, Subramanian, Santos, Mehri, Rostamzadeh, Bengio, and Pal]{trabelsi2017deep}
Chiheb Trabelsi, Olexa Bilaniuk, Ying Zhang, Dmitriy Serdyuk, Sandeep Subramanian, Joao~Felipe Santos, Soroush Mehri, Negar Rostamzadeh, Yoshua Bengio, and Christopher~J Pal.
\newblock Deep complex networks.
\newblock \emph{arXiv preprint arXiv:1705.09792}, 2017.

\bibitem[Bassey et~al.(2021)Bassey, Qian, and Li]{bassey2021survey}
Joshua Bassey, Lijun Qian, and Xianfang Li.
\newblock A survey of complex-valued neural networks.
\newblock \emph{arXiv preprint arXiv:2101.12249}, 2021.

\bibitem[Lee et~al.(2022)Lee, Hasegawa, and Gao]{lee2022complex}
ChiYan Lee, Hideyuki Hasegawa, and Shangce Gao.
\newblock Complex-valued neural networks: A comprehensive survey.
\newblock \emph{IEEE/CAA Journal of Automatica Sinica}, 9\penalty0 (8):\penalty0 1406--1426, 2022.

\bibitem[Axelsson and Kucherov(2000)]{axelsson2000real}
Owe Axelsson and Andrey Kucherov.
\newblock Real valued iterative methods for solving complex symmetric linear systems.
\newblock \emph{Numerical linear algebra with applications}, 7\penalty0 (4):\penalty0 197--218, 2000.

\bibitem[Benzi and Bertaccini(2008)]{benzi2008block}
Michele Benzi and Daniele Bertaccini.
\newblock Block preconditioning of real-valued iterative algorithms for complex linear systems.
\newblock \emph{IMA Journal of Numerical Analysis}, 28\penalty0 (3):\penalty0 598--618, 2008.

\bibitem[Day and Heroux(2001)]{day2001solving}
David Day and Michael~A Heroux.
\newblock Solving complex-valued linear systems via equivalent real formulations.
\newblock \emph{SIAM Journal on Scientific Computing}, 23\penalty0 (2):\penalty0 480--498, 2001.

\bibitem[Saad and Schultz(1986)]{saad1986gmres}
Youcef Saad and Martin~H Schultz.
\newblock Gmres: A generalized minimal residual algorithm for solving nonsymmetric linear systems.
\newblock \emph{SIAM Journal on scientific and statistical computing}, 7\penalty0 (3):\penalty0 856--869, 1986.

\bibitem[Gutknecht(1993)]{gutknecht1993variants}
Martin~H Gutknecht.
\newblock Variants of bicgstab for matrices with complex spectrum.
\newblock \emph{SIAM journal on scientific computing}, 14\penalty0 (5):\penalty0 1020--1033, 1993.

\bibitem[Freund(1992)]{freund1992conjugate}
Roland~W Freund.
\newblock Conjugate gradient-type methods for linear systems with complex symmetric coefficient matrices.
\newblock \emph{SIAM Journal on Scientific and Statistical Computing}, 13\penalty0 (1):\penalty0 425--448, 1992.

\bibitem[Hoffreumon and Woods(2025)]{hoffreumon2025quantum}
Timothee Hoffreumon and Mischa~P Woods.
\newblock Quantum theory does not need complex numbers.
\newblock \emph{arXiv preprint arXiv:2504.02808}, 2025.

\bibitem[Erd{\"o}s and Moser(1964)]{erdos1964problem}
P~Erd{\"o}s and Leo Moser.
\newblock A problem on tournaments.
\newblock \emph{Canadian Mathematical Bulletin}, 7\penalty0 (3):\penalty0 351--356, 1964.

\bibitem[Moon et~al.(1968)]{moon1968topics}
John~W Moon et~al.
\newblock Topics on tournaments.
\newblock \emph{New York}, 1968.

\bibitem[Lu et~al.(2016)Lu, Biamonte, Li, Li, Johnson, Bergholm, Faccin, Zimbor{\'a}s, Laflamme, Baugh, et~al.]{lu2016chiral}
Dawei Lu, Jacob~D Biamonte, Jun Li, Hang Li, Tomi~H Johnson, Ville Bergholm, Mauro Faccin, Zolt{\'a}n Zimbor{\'a}s, Raymond Laflamme, Jonathan Baugh, et~al.
\newblock Chiral quantum walks.
\newblock \emph{Physical Review A}, 93\penalty0 (4):\penalty0 042302, 2016.

\bibitem[Golub and Van~Loan(2013)]{golub2013matrix}
Gene~H. Golub and Charles~F. Van~Loan.
\newblock \emph{Matrix Computations}.
\newblock JHU Press, 2013.

\bibitem[Reichel and Trefethen(1992)]{reichel1992eigenvalues}
Lothar Reichel and Lloyd~N Trefethen.
\newblock Eigenvalues and pseudo-eigenvalues of toeplitz matrices.
\newblock \emph{Linear algebra and its applications}, 162:\penalty0 153--185, 1992.

\bibitem[Gray et~al.(2006)]{gray2006toeplitz}
Robert~M Gray et~al.
\newblock Toeplitz and circulant matrices: A review.
\newblock \emph{Foundations and Trends{\textregistered} in Communications and Information Theory}, 2\penalty0 (3):\penalty0 155--239, 2006.

\bibitem[Daskin and Kais(2011)]{daskin2011decomposition}
Anmer Daskin and Sabre Kais.
\newblock Decomposition of unitary matrices for finding quantum circuits: application to molecular hamiltonians.
\newblock \emph{The Journal of chemical physics}, 134\penalty0 (14), 2011.

\bibitem[Javadi-Abhari et~al.(2024)Javadi-Abhari, Treinish, Krsulich, Wood, Lishman, Gacon, Martiel, Nation, Bishop, Cross, Johnson, and Gambetta]{qiskit2024}
Ali Javadi-Abhari, Matthew Treinish, Kevin Krsulich, Christopher~J. Wood, Jake Lishman, Julien Gacon, Simon Martiel, Paul~D. Nation, Lev~S. Bishop, Andrew~W. Cross, Blake~R. Johnson, and Jay~M. Gambetta.
\newblock Quantum computing with {Q}iskit, 2024.

\bibitem[Virtanen et~al.(2020)Virtanen, Gommers, Oliphant, Haberland, Reddy, Cournapeau, Burovski, Peterson, Weckesser, Bright, {van der Walt}, Brett, Wilson, Millman, Mayorov, Nelson, Jones, Kern, Larson, Carey, Polat, Feng, Moore, {VanderPlas}, Laxalde, Perktold, Cimrman, Henriksen, Quintero, Harris, Archibald, Ribeiro, Pedregosa, {van Mulbregt}, and {SciPy 1.0 Contributors}]{2020SciPy-NMeth}
Pauli Virtanen, Ralf Gommers, Travis~E. Oliphant, Matt Haberland, Tyler Reddy, David Cournapeau, Evgeni Burovski, Pearu Peterson, Warren Weckesser, Jonathan Bright, St{\'e}fan~J. {van der Walt}, Matthew Brett, Joshua Wilson, K.~Jarrod Millman, Nikolay Mayorov, Andrew R.~J. Nelson, Eric Jones, Robert Kern, Eric Larson, C~J Carey, {\.I}lhan Polat, Yu~Feng, Eric~W. Moore, Jake {VanderPlas}, Denis Laxalde, Josef Perktold, Robert Cimrman, Ian Henriksen, E.~A. Quintero, Charles~R. Harris, Anne~M. Archibald, Ant{\^o}nio~H. Ribeiro, Fabian Pedregosa, Paul {van Mulbregt}, and {SciPy 1.0 Contributors}.
\newblock {{SciPy} 1.0: Fundamental Algorithms for Scientific Computing in Python}.
\newblock \emph{Nature Methods}, 17:\penalty0 261--272, 2020.
\newblock \doi{10.1038/s41592-019-0686-2}.

\bibitem[Higham(2005)]{higham2005scaling}
Nicholas~J. Higham.
\newblock The scaling and squaring method for the matrix exponential revisited.
\newblock \emph{SIAM Review}, 47\penalty0 (1):\penalty0 3--49, 2005.

\bibitem[Moler and Van~Loan(2003)]{moler2003nineteen}
Cleve Moler and Charles Van~Loan.
\newblock Nineteen dubious ways to compute the exponential of a matrix, twenty-five years later.
\newblock \emph{SIAM Review}, 45\penalty0 (1):\penalty0 3--49, 2003.

\bibitem[Shende et~al.(2005)Shende, Bullock, and Markov]{shende2005synthesis}
Vivek~V Shende, Stephen~S Bullock, and Igor~L Markov.
\newblock Synthesis of quantum logic circuits.
\newblock In \emph{Proceedings of the 2005 Asia and South Pacific Design Automation Conference}, pages 272--275, 2005.

\end{thebibliography}
\end{document}